\documentclass[twocolumn]{aastex63}

\usepackage{amsmath}
\usepackage{float}

\usepackage{ulem}

\usepackage{graphicx}
\usepackage{bm}
\usepackage{times}
\usepackage{url}
\usepackage{wasysym}
\usepackage{xcolor}
\usepackage[switch]{lineno}
\usepackage{hyperref}

\makeatletter
\let\tablenum\@undefined
\makeatother
\usepackage{siunitx}

\newcommand{\pdiff}[2]{\frac{\partial #1}{\partial #2}}

\DeclareSIUnit\gauss{G}
\DeclareSIUnit\erg{erg}
\DeclareSIUnit\pc{pc}
\DeclareSIUnit\year{yrs}

\received{XXX, 2020}
\revised{yyy, 2020}
\accepted{zzz, 2020}
\submitjournal{ApJ}

\shorttitle{Ultra-high energy Inverse Compton emission from Galactic electron accelerators}
\shortauthors{Breuhaus et al.}

\graphicspath{{./}{figures/}}

\begin{document}

\title{Ultra-high energy Inverse Compton emission from Galactic electron accelerators}


\correspondingauthor{Mischa Breuhaus}
\email{mischa.breuhaus@mpi-hd.mpg.de}

\author[0000-0003-0268-5122]{M. Breuhaus}
 \author{J. Hahn}
 \altaffiliation{Now at: CGI Darmstadt, Rheinstra{\ss}e 95, 64295 Darmstadt, Germany}
 \author[0000-0003-2541-4499]{C. Romoli}
 \author[0000-0002-3778-1432]{B. Reville}
 \author[0000-0001-9745-5738]{G. Giacinti}
 \author{R. Tuffs}
 \author[0000-0002-1031-7760]{J.~A. Hinton} 
\affiliation{Max-Planck-Institut f\"ur Kernphysik, Saupfercheckweg 1, 69117 Heidelberg, Germany}

\begin{abstract}
It is generally held that $>$100~TeV emission from astrophysical objects unambiguously demonstrates the presence of PeV protons or nuclei, due to the unavoidable Klein-Nishina suppression of inverse Compton emission from electrons. However, in the presence of inverse Compton dominated cooling, hard high-energy electron spectra are possible. We show that the environmental requirements for such spectra can naturally be met in spiral arms, and in particular in regions of enhanced star formation activity, the natural locations for the most promising electron accelerators: powerful young pulsars. Our scenario suggests a population of hard ultra-high energy sources is likely to be revealed in future searches, and may also provide a natural explanation for the 100~TeV sources recently reported by HAWC.

\end{abstract}


\section{Introduction} \label{sec:00}

The search for sources of cosmic rays at energies above $10^{15}$~eV remains one of the key challenges in high-energy astrophysics. Gamma-ray instruments that survey the $>$100~TeV sky are at last close to resolving this century old question. 
The HAWC collaboration recently reported the discovery of significant emission extending beyond 100~TeV from a number of Galactic sources~\citep{HAWC_UHE}. At first sight the detection of
hard-spectrum $\gamma$-ray emission up to these extreme energies looks like clear evidence of the acceleration of protons and nuclei, rather than Inverse Compton (IC) emission, due to the inevitable Klein–Nishina (KN) suppression 
at such energies. However, in specific circumstances the emission of accelerated electrons may produce spectra compatible with these observations: in particular when electron cooling is dominated by IC losses~\citep{BG70, Zdziarski93, HintonAharonian}. The exploration of this scenario is particularly important as the reported HAWC ultra-high-energy (UHE) sources exhibit no obvious correlation with target material as would be expected in the hadronic scenario, and all are associated with young and powerful pulsars. The nebulae associated to such pulsars are well-established TeV-emitters with strong evidence that this emission has a dominant IC origin~\cite[e.g.][]{HESSJ1825}.

A detailed understanding of such sources of high-energy electrons and positrons is significant also for other fields. For example, the local positron flux \citep{PAMELA,AMS} is thought to be dominated by nearby pulsars \cite[e.g.][]{EDGE}, however, possible dark matter annihilation signatures \citep{Bergstrom, Cholis} cannot yet be ruled out. Knowledge of the astrophysical background is paramount in this effort.
Similarly, UHE sources are powerful tools in the search for new physics, including searches for violation of Lorentz Invariance~\citep{Martinez-Huerta2017}. 
An understanding of source properties is essential to constrain such theories.

In the following, we explore the emission from high-energy electrons/positrons in radiation-dominated environments. 
In section \ref{sec:01}, the effect of the photon field properties on the $\gamma$-ray spectra are reviewed. Section \ref{sec:02} outlines the requirements on accelerators' surroundings for hard spectra to be observed at UHE. The potential for favourable conditions on Galactic scales are discussed in section \ref{sec:03}, where it is shown that active star-forming regions (SFRs) are likely to satisfy the necessary environmental requirements.
The model is tested against selected UHE HAWC source spectra in section \ref{sec:04}. Conclusions are presented in section \ref{sec:05}.

\begin{figure*}
  \begin{center}
  \includegraphics[height=0.393\hsize]{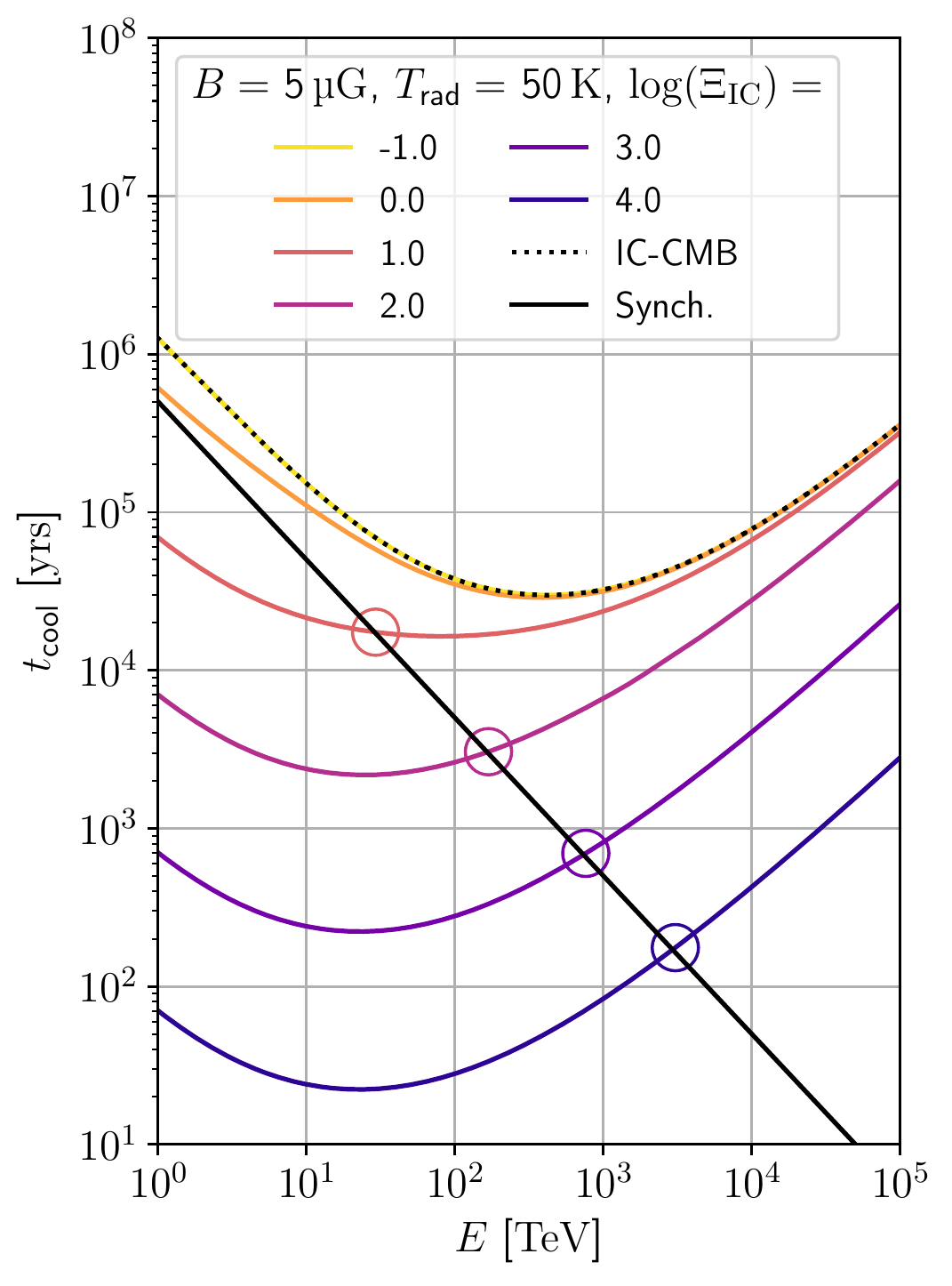}
  \includegraphics[height=0.393\hsize]{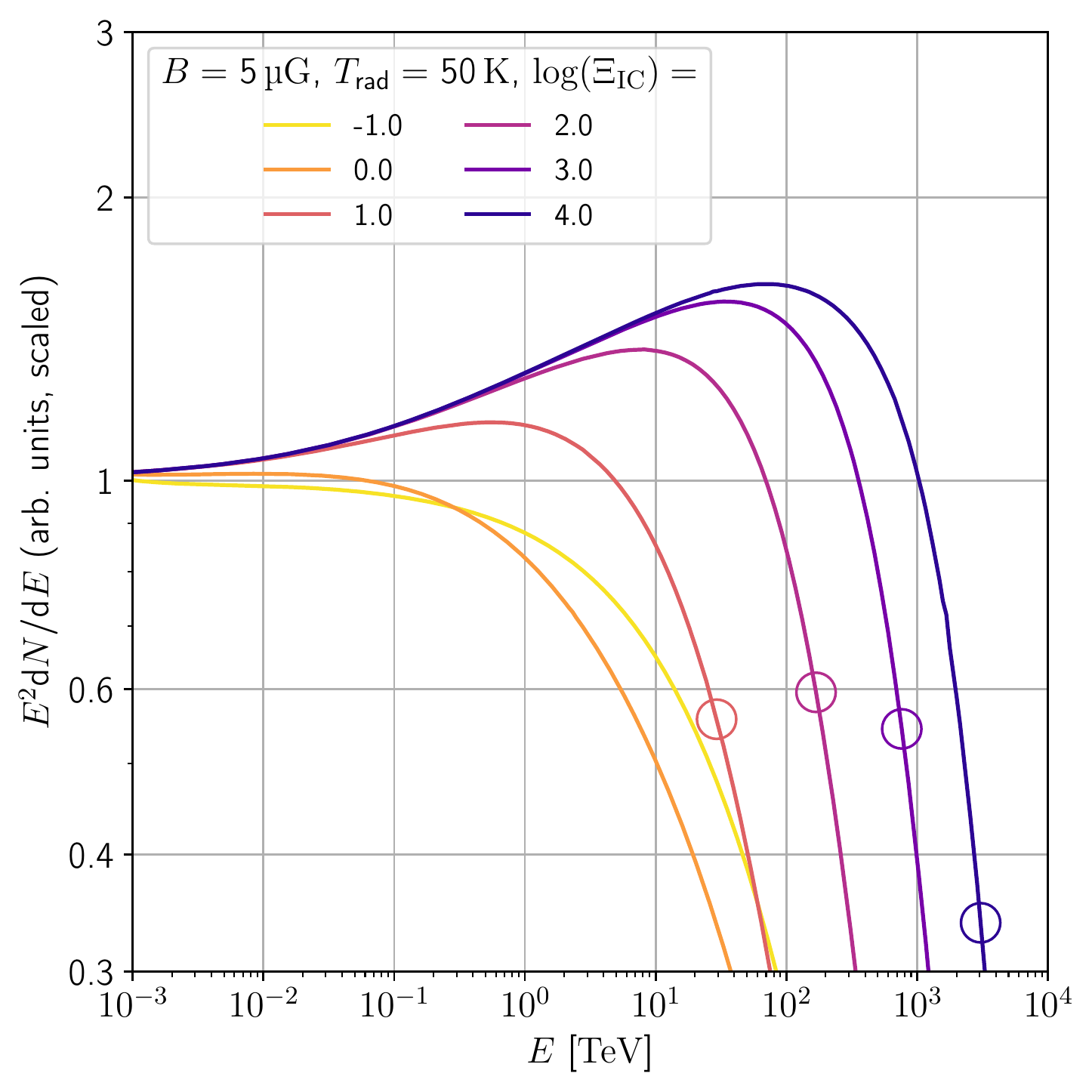}
  \includegraphics[height=0.393\hsize]{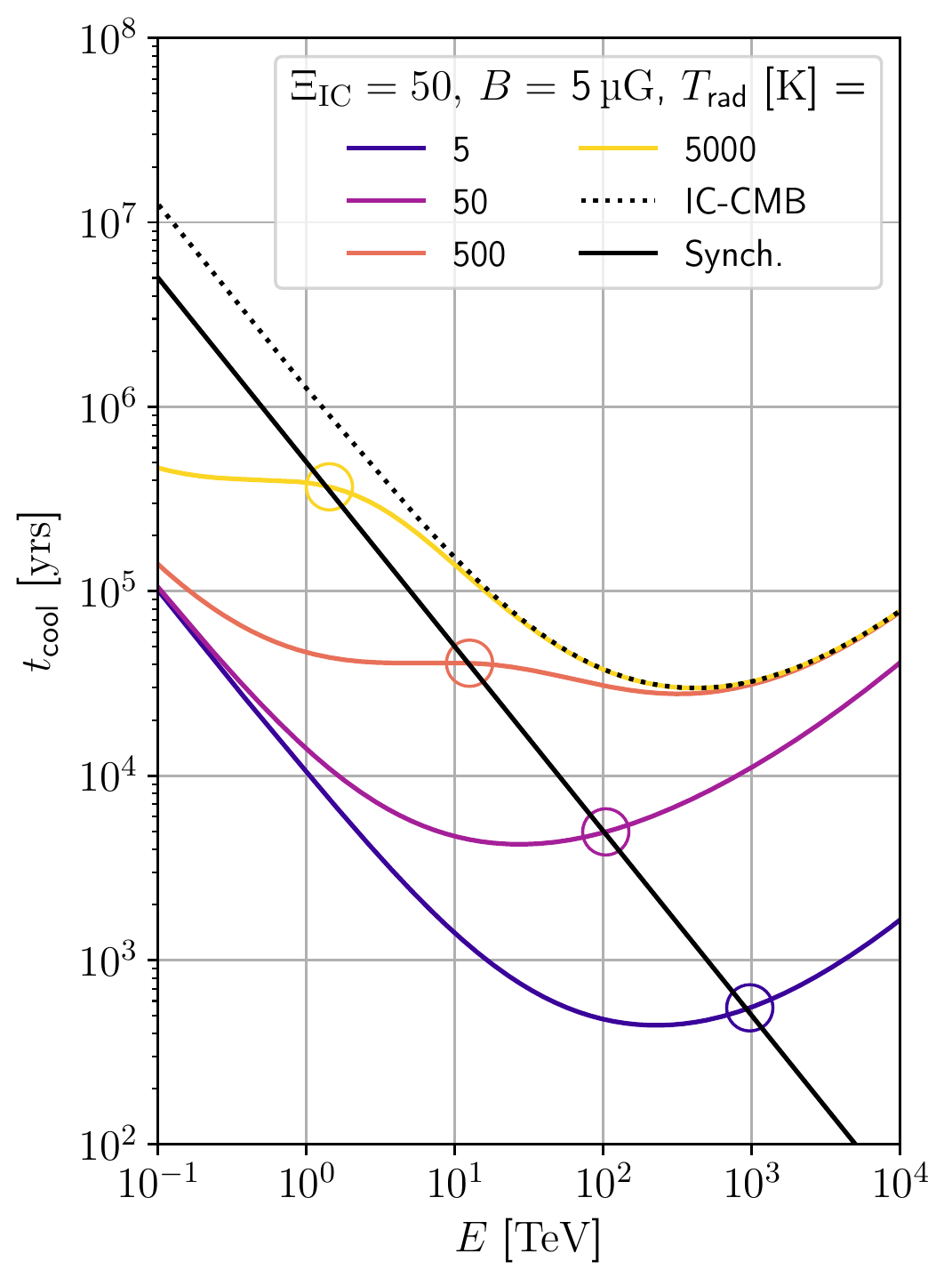}
\end{center}
  \caption{Electron cooling time-scales $t_{\text{cool}}$ (left) and the steady-state $\gamma$-ray (middle) spectra assuming different
 energy densities of a 50~K radiation field including the CMB, for a representative fixed magnetic field of 5~$\mu$G. The $\gamma$-ray spectra are those arising in equilibrium from continuous injection of an $E^{-2}$ spectrum with 
 exponential cutoff at \SI{10}{\peta\electronvolt}. The right panel shows $t_{\text{cool}}$ for fixed $\Xi_{\rm IC}=50$, but different temperatures.
The circles on all panels indicate the transition energy $E_{X}$ for each radiation field density. Spectra for more sophisticated galactic radiation fields are provided in Appendix \ref{sec:AppendixA}.}
  \label{fig:break}
\end{figure*}

\section{Inverse-Compton spectra in radiation dominated environments } \label{sec:01}

UHE electrons in astrophysical environments have short cooling times, with energy losses dominated by synchrotron and IC emission. The cooling time in the Thomson regime is
$$
t_{\rm cool}\,\approx\,300\,\left(\frac{E_{e}}{\rm 1\,PeV}\right)^{-1}\left(\frac{U_{B}+U_{\rm rad}}{ {\rm 1\,eV\,cm^{-3}}}\right)^{-1} \mbox{years}\, , \enspace\enspace
$$
\noindent
where $U_{B}$ and $U_{\rm rad}$ are the magnetic and radiation field energy densities respectively. 
Such rapid cooling means that the electron spectrum at these energies can be assumed to result in equilibrium between injection/acceleration and losses. 
If the magnetic energy density exceeds that of the radiation energy density, i.e. $\Xi_{\rm IC} \equiv U_{\rm rad}/U_{B} \ll 1$, a region with continuous power-law injection $(\propto E^{-\alpha})$ leads to $dN/dE \propto E^{-(\alpha+1)}$ spectrum in equilibrium. The resulting $\gamma$-ray emission is a (broken) power-law with photon-index $\Gamma = -(\alpha+2)/2$ in the Thomson regime, softening to $-(\alpha+2)$ in the KN regime ($E_{\rm KN} \sim m_e^{2}c^{4}/E_{\rm rad}$, where $E_{\rm rad}$ is the target photon energy.).
The situation is different when $\Xi_{\rm IC} \gg 1$ where the energy dependence of the KN cross-section leads to a hardening of the equilibrium electron spectrum~\citep{BG70, Moderski2005}. This hardening is less pronounced in the resulting $\gamma$-ray spectrum, due to KN suppression of the emission. Crucially in radiation-dominated environments it is possible to maintain hard $\gamma$-ray spectra well beyond $E_{\rm KN}$. Above a critical energy synchrotron losses inevitably dominate.
We define the cross-over energy, $E_{X}$ as the electron energy at which the synchrotron cooling time equals that of IC. To explore the environmental dependence of IC spectra we use the {\it GAMERA} code~\citep{gamera}. Figure~\ref{fig:break} shows how $E_{X}$ is related to the resulting $\gamma$-ray spectrum (left and middle panel). As $E_{X}$ must occur in the KN regime, electrons lose most of their energy during a single scattering process and hence the feature in the $\gamma$-ray spectrum occurs at essentially the same energy ($E_{X}$) as in the electron spectrum. For sufficiently large $\Xi_{\rm IC}$, a hard $\gamma$-ray spectrum is produced. The hardening is a consequence of the increased cooling-time for energies $E_{\rm KN}<E<E_{X}$ compensating for the reduced efficiency in IC scattering \citep{Zdziarski}. This feature is easily masked in a realistic source, for example by a softening or cut-off in the injected electron spectrum.

In the following, we focus primarily on the effect of strong infra-red (IR) fields, being the most important for UHE emission, 
although the cosmic microwave background (CMB) may also play a role. 
While the CMB alone is not likely to produce spectral-hardening, it does provide the dominant target photon population after KN effects suppress IR scattering. 
This is evident from comparison of the shapes for the two curves: $\Xi_{\rm IC} = 1 ~\&~0.1$ in the middle panel of Figure \ref{fig:break}. Note that the fluxes are normalised to match at low energies to highlight the important physical features, the spectral shape near the cut-off in this case.

The temperature of the IR field also impacts the IC spectrum. For the $5 \mu$G magnetic field and 50~K black-body photon field adopted in the left \& middle panels of Figure~\ref{fig:break} a $\sim100$ TeV IC emitter requires $\Xi_{\rm IC}>10$.
The right panel of Figure~\ref{fig:break} shows the $E_{X}$ dependence on black-body temperature, for fixed $\Xi_{\rm IC}=50$. Lower temperatures are advantageous to move $E_{X}$, and the resulting IC emission toward the UHE range.

Figure~\ref{fig:par_space} summarises the available phase-space for hard spectrum $\gamma$-ray emission from electron accelerators in terms of radiation dominance level and ambient photon field temperature, again for a \SI{5}{\micro\gauss}magnetic field. For the CMB the required $\Xi_{\rm IC}$ is \num{\sim 3}, implying a magnetic field of \SI{1.8}{\micro\gauss}. Low $B$ field values reduce the required values of $\Xi_{\rm IC}$ at all temperatures because of the increasing influence of the CMB.

Realistic Galactic photon fields will be different than the simplified case of the CMB and black-body IR spectrum. The main effect above \si{\tera\electronvolt} energies is a slightly softer $\gamma$-ray spectrum because parts of the radiation field are in the KN-regime. A small decrease in the power-law index $\alpha$ of the injection electron spectrum will compensate this effect and the characteristics of the resulting $\gamma$-ray spectrum are the same as the ones for the models in Figure \ref{fig:break}, where $T$ is the characteristic temperature of the FIR emission, see Appendix \ref{sec:AppendixA}.
In the calculations above, electrons are injected with a sufficiently high cut-off energy and the $\gamma$-rays reach the observer without losses. This requires both efficient acceleration and low absorption, which imposes additional constraints on the system. We explore these, and other limits in the next section.

\section{System Constraints} \label{sec:02}

The requirement of large $\Xi_{\rm IC}$ imposes additional constraints. Acceleration of particles to energies in excess of $100$ TeV requires the magnetic field to be strong enough to confine the particles close to the accelerator \citep{Hillas1984}. 
If the size of the accelerator is significantly smaller than for example the $100$ TeV HAWC sources (see Figure \ref{fig:hillas_and_gammagamma_limits}), the resulting limits on magnetic energy density places severe demands on the photon energy density.   
However, as we show below, if the magnetic field decreases with distance from the acceleration site, the field strength there is not critical. For a homogeneous soft photon field, large $\Xi_{\rm IC}$ can therefore naturally emerge at larger radii.
On the other hand, large path-lengths through a radiation dominated environment may lead to significant $\gamma\gamma$ pair-production. In addition, triplet pair-production should be considered in the deep KN regime. We explore these aspects in turn.

\paragraph{I Acceleration requirements} Pulsars with spin-down luminosity $L_{\rm SD} >10^{36}~{\rm erg /s}$ are potential PeV electron accelerators \cite[e.g.][]{Hillas1984}. For the Crab pulsar theoretical and observational evidence supports this conclusion \citep{BuehlerBlandford}.
As mentioned previously, requiring $\Xi_{\rm IC} \gg 1$ while maintaining equipartition between the spin-down luminosity and magnetic energy flux suggests the acceleration should occur far from the central pulsar where the magnetic field is weak: $B_{\rm eq} \approx 3 (L_{\rm SD}/10^{36} {\rm erg~s}^{-1})^{1/2} (r/{\rm pc})^{-1}~\mu$G. It is still possible to generate hard equilibrium spectra when the acceleration zone has $\Xi_{\rm IC} \ll 1$ provided accelerated particles can diffuse away from the acceleration site before suffering significant losses. Such a scenario is likely if the field decreases with radius such as is found in the high magnetization post-shock solutions of \citet{KennelCoroniti} where the magnetic field decreases as $1/r$. The synchrotron cooling time thus increases with radius, and hence the homogeneous soft-photon field will dominate at sufficiently large radii.
The position of the cut-off in the electron spectrum in such a scenario depends on how quickly particles diffuse away from the acceleration site.
To quantify the impact of synchrotron losses on the maximum energy when the cooling time has an $r^2$ dependence, the spherically symmetric radial transport equation is solved. 
The steady-state solution is given in Appendix \ref{sec:AppendixC}. Assuming a fraction $\eta_{\rm eq}$ of the spin-down luminosity is carried by magnetic flux, a cut-off in the spectrum due to synchrotron cooling alone can be shown to occur at
$E_{\rm br} \approx 0.5 (\eta_{\rm eq} L_{36})^{-1} D_{26}~ {\rm PeV}$, where $L_{\rm SD} = L_{36}10^{36}~{\rm erg /s}$ and $D=10^{26} D_{26}$ cm$^2$/s is the spatial diffusion coefficient.  
Thus, even in cases where the acceleration site is magnetically dominated,  high-power pulsars can in principle feed a large $\Xi_{\rm IC} > 1$ volume with $>100$ TeV electrons.

\begin{figure}
	\begin{center}
		\hspace*{-0.3cm}\resizebox{1.0\hsize}{!}{\includegraphics[clip=]{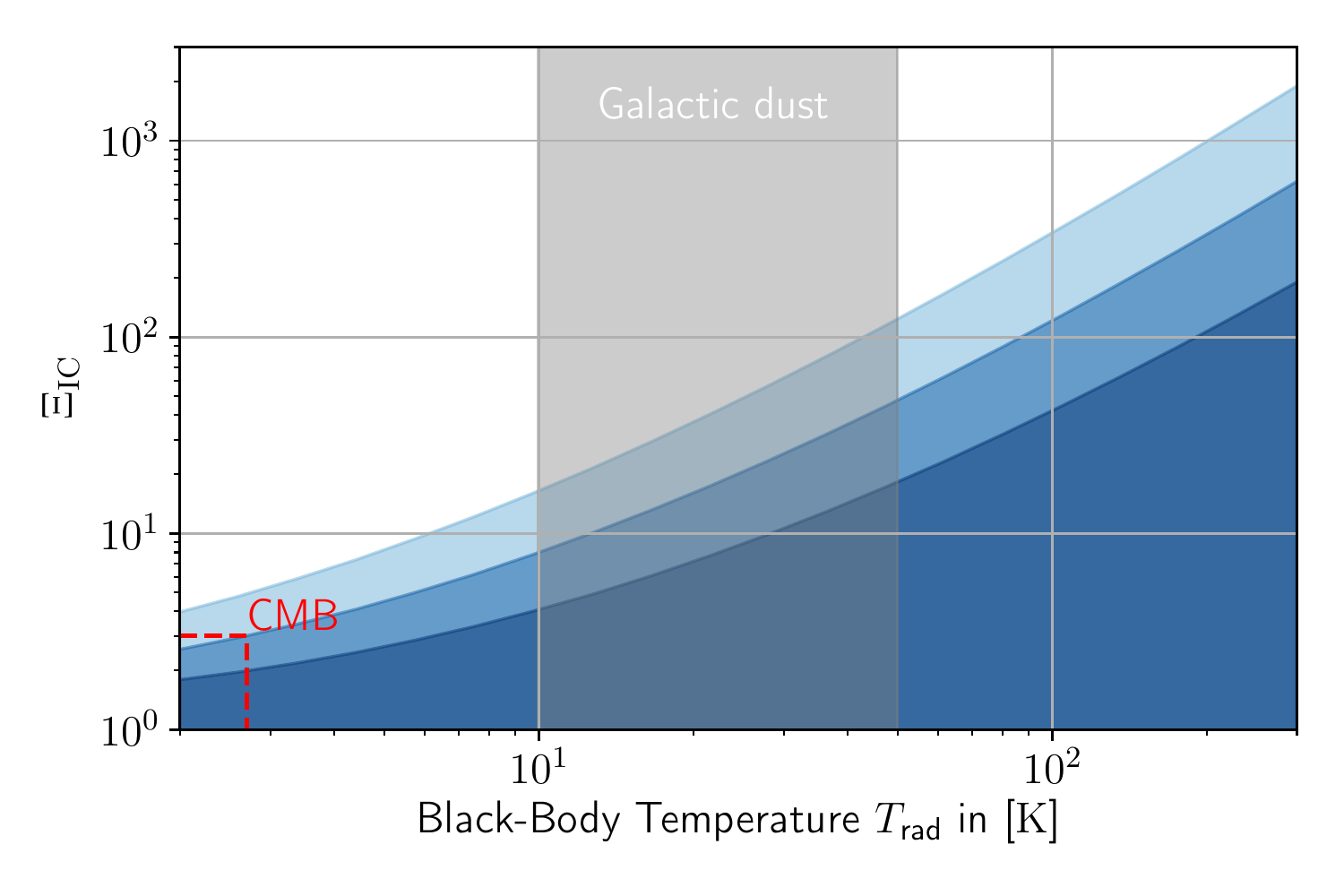}}
		\caption{The blue shaded areas show the regions of $T_{\rm rad}$-$\Xi_{\rm IC}$ space in which values of $E_{\rm X}$ above a given specified energy are excluded. Larger values of $E_{\rm X}$ are necessary to allow hard IC spectra. From bottom up the exclusion zones correspond to $E_{X}=50,100,200$~TeV. The ratio for $E_{X}=\SI{100}{\tera\electronvolt}$ for the CMB {temperature} is marked in red and the vertical grey band highlights the typical dust-temperature range from 10--50~K~\citep{Bernard_et_al_2010, Zhu_Huang_2014}. {The $B$-field was taken to be \SI{5}{\micro\gauss} and the CMB was added to the black body photon field. For lower $B$-fields, lower values of $\Xi_{\rm IC}$ are required, because of the increasing influence of the CMB.}}
		\label{fig:par_space}
	\end{center}
\end{figure}

\paragraph{II Opacity requirements} The high radiation-field environment needed for UHE IC sources may also lead to significant $\gamma\gamma$ pair production. Peak absorption occurs when $E_\gamma \approx 3.5 E_{\rm KN}$ \citep{gammagamma}. The wavelength corresponding to peak absorption is  $\lambda_{\rm peak} \approx\,136\,(E_{\gamma}/100~{\rm TeV})\,\mu$m. Far infrared (FIR) radiation is therefore critical and can limit detection of UHE photons.
For sources, the level of absorption depends on the photon field details, as well as the
path-length of the emerging $\gamma$-rays through the surroundings. While stronger FIR fields give larger values of
$E_{X}$, the resulting attenuation of $\gamma$-rays sets a limit on the maximum source size.

The combined acceleration and pair-production limits are shown in Figure~\ref{fig:hillas_and_gammagamma_limits}. The apparent sizes of the $100$\,TeV HAWC sources are also indicated. The Hillas limit is calculated for PeV electrons, and represents a minimal requirement for confinement. The black-body temperatures coincide with Galactic dust boundaries (see Figure \ref{fig:par_space}). The $\gamma\gamma$ absorption upper limits correspond to $1/e$ attenuation for $100$~TeV photons. The normalisation of the absorption curves is chosen for a FIR field with fixed $\Xi_{\rm IC}$. For the $T_{\rm rad} = $\SIlist{50(10)}{\kelvin} curve, we set $\Xi_{\rm IC}=55(8.8)$ such that $E_{\text X}$ is at least $100$\,TeV at all field strengths. In lower magnetic fields, the CMB becomes increasingly important, and the value of $E_{\text X}$ for a given $\Xi_{\rm IC}$ increases. At higher $\gamma$-ray energies, the attenuation length decreases, and above \SI{\sim 300}{\tera\electronvolt}, $\gamma\gamma$ absorption is dominated by CMB photons \citep{2016PhRvD..94f3009V, Popescu2017}. Finally we consider the attenuation of 100~TeV photons on larger scales using an axisymmetric radiation model of the Milky Way \citep{Popescu2017}. We find that even in the worst-case scenario of sources located at the opposite edge of the Galaxy, transmission factors for 100~TeV photons are $e^{-\tau_{\gamma\gamma}}>0.5$, where $\tau_{\gamma\gamma}$ is the optical depth.

\begin{figure}

	\centering
	\hspace*{-0.1cm}\resizebox{1.0\hsize}{!}{\includegraphics[clip=]{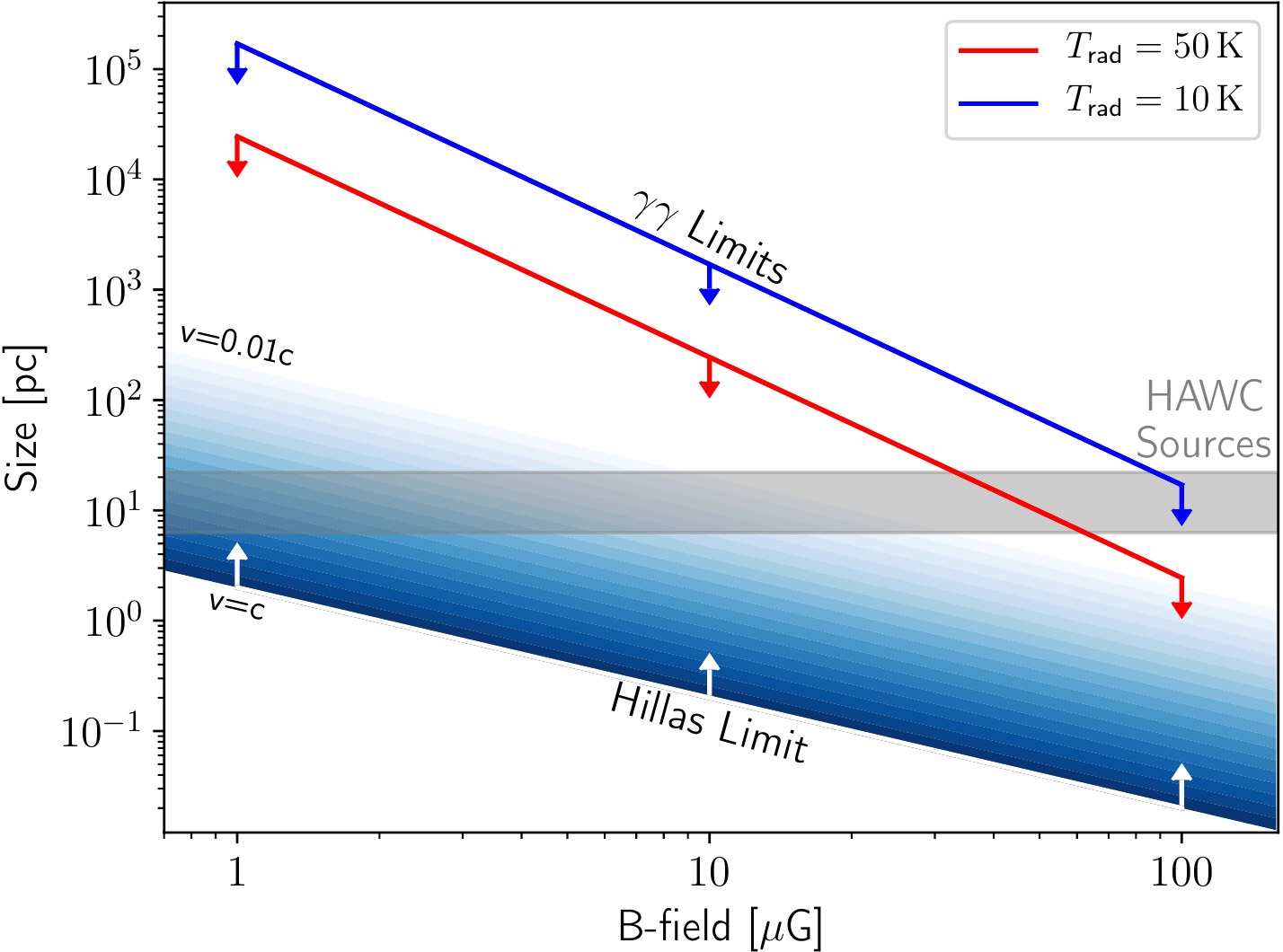}}
	\caption{Limits on the linear size of UHE IC sources due to confinement/acceleration and $\gamma\gamma$ pair production. The Hillas limits correspond to an electron energy of \SI{1}{\peta\electronvolt}. The grey horizontal band shows the extent of the three \SI{100}{\tera\electronvolt} sources reported by HAWC \citep{HAWC_UHE}.
		The $\gamma\gamma$-absorption limits give the $1/e$ attenuation length of \SI{100}{\tera\electronvolt} photons
		with a fixed value of $\Xi_{\rm IC} = 55(8.8)$ for $T_{\text{rad}} = {50(10)}$K. The values are chosen such that
		$E_{X}=100$ TeV at high magnetic-field strengths. 
	}
	\label{fig:hillas_and_gammagamma_limits}
\end{figure}

\paragraph{III Triplet Pair Production} In extreme cases, the triplet pair production (TPP) process, $\gamma e \rightarrow 3 e$, should also be considered. The cross section for TPP exceeds that of KN for collision energies above $x=E E_{\rm rad}/m_{\rm e}^2 c^4 \approx 250$ \cite[e.g.][]{MMB}. The losses per collision due to TPP are more gradual than IC scattering in deep KN, since the daughter pair acquire only a small fraction of the scattered electron's energy. While the TPP cooling-rate does not dominate until $x\approx (4 \alpha_{\rm f})^{-4}\approx 10^6$, the effect of losses due to TPP are already noticeable above $x\approx 10^4$ \citep{Mastichiadis91}. For $100$ TeV electrons, the latter condition corresponds to $E_{\rm rad}>25$ eV. Thus if the photon energy density is concentrated in ionizing UV or above, spectral hardening might be less pronounced. Note however, that for TPP to dominate over synchrotron losses requires $\Xi_{\rm IC} \sim 10^9$ \citep{DS91}. A more serious consequence of TPP regarding the generation of hard spectra is the inevitable IC emission of the secondary pairs produced, which emit lower energy photons, steepening the resulting equilibrium $\gamma$-ray spectrum. As discussed by \citet{Mastichiadis91}, the effect only manifests for hard electron-injection spectra $\alpha < 2$. Numerical simulations reveal that, in the absence of synchrotron losses, for hard injection spectra with $\alpha < 1.75$ the equilibrium $\gamma$-ray photon spectrum asymptotes to $dN_{\gamma}/dE \propto E^{-1.75}$. We conclude that while TPP does not prevent the production of hard $\gamma$-ray spectra in the UHE regime, it should be accounted for in any detailed modelling of individual sources.

\section{Environmental Constraints}\label{sec:03}

 Having established the criteria for hard spectrum UHE IC emitters, we now assess if these conditions can be met within our galaxy. On large scales, models exist for both radiation densities and magnetic fields in the Milky Way. As a large scale radiation field model we adopt that of \citet{Popescu2017} including CMB. Magnetic fields are in general less well constrained. As a representative model we consider
 \citet{Jansson2012,Jansson2012_2}. Combining these to derive $\Xi_{\rm IC}$ values throughout the Galaxy, promising locations ($\Xi_{\rm IC} \gg 1$) are found only outside of spiral arms and/or the Galactic disk. As sources are less likely to be found in such regions we conclude that, despite uncertainties, conditions in typical very large scale environments are unlikely to be suitable for UHE IC emission. 
 
The situation may be different in local regions of active star formation. There is a clear general association of known TeV $\gamma$-ray sources to star-forming activity, both via the established TeV source classes and the overall spatial distribution of TeV sources in our Galaxy~\citep{hessscanpaper, hgps}. Regions of intense star-formation activity are likely to exhibit conditions distinctly different from the general Galactic environment, particularly in terms of radiation fields, due to emission from clusters of massive stars and emission arising from the heating of the local environment by this intense radiation. As noted earlier, FIR radiation fields are the most likely to result in hard spectrum IC emission, and intense FIR emission is a characteristic of young and compact star-forming regions where the UV emission of massive stars is reprocessed by dust.

Models of the evolution of dusty photo-dissociation regions bounding massive stellar clusters are
typically consistent with an incident non-ionising UV radiation field of several hundred eV\,cm$^{-3}$ 
maintaining a local FIR energy density up to $\sim$100~eV\,cm$^{-3}$ over a substantial volume within the star-forming region \citep{Dopitaetal2005,Grovesetal2008,Popescu2011}.
This will be dependent on the age and mass of the star clusters and the geometrical distribution of the gas clouds. 
We conclude that conditions in massive stellar clusters are ideal for hard UHE sources. While the HAWC UHE source locations are not remarkable in terms of association to obvious regions of ongoing star formation \citep{Murray_Rahman2010}, as we show below, they may still be sampling atypical conditions. 

Taking the UHE HAWC sources as examples, based on assumed proximity to their associated high-power pulsars ($\sim$5--8\,kpc) and galactocentric radii, the
axisymmetric model of \citet{Popescu2017} gives IR energy densities ranging from 0.3--1.2~eV\,cm$^{-3}$. 
However, local enhancements in the radiation field density are to be expected inside spiral arms, with respect to the axisymmetric radiation model, roughly proportional to the inverse of the volume filling factor of spiral arms in the disk. Variations in FIR density within spiral arms are also inevitable, due to fluctuations in local star-formation activity and reprocessing by dust, even for sources not in the immediate vicinity of stellar clusters. Significant complexity is already revealed in the surroundings of the UHE HAWC sources \citep{Voisin,Paredes09}. 

Just as enhanced pockets of diffuse FIR emission are expected in spiral arms, the large-scale $B$-field is also increased with respect to that in the inter-arm regions. However, the largest contribution to the absolute strength is expected to be due to local random field components and not the large-scale regular field. Therefore, strong fluctuations on scales of \SI{100}{\pc} or less will occur and regions with $B \lesssim \SI{3}{\micro\gauss}$ are possible. Such low magnetic energy densities may also occur in the interiors of superbubbles \citep{Korpi_et_al_1999}, or more locally to the pulsar wind nebulae if electrons are channelled into an adjacent low-density region \citep{PWNHalos}. 
As enhanced FIR emission regions, superbubbles, and the presence of powerful pulsars are all associated to previous or ongoing star-formation activity, a spatial coincidence seems highly likely.
Suitable environments may therefore exist along essentially any line of sight through the inner galaxy.

\section{Application to UHE HAWC sources}
\label{sec:04}

Although none of the UHE HAWC sources are associated to specific SFRs, maps from the Infrared Astronomical Satellite (IRAS) \citep{IRAS} reveal multiple discrete sources nearby, which indicate enhancements relative to the axisymmetric diffuse emission.
We calculate the additional contributions from IRAS discrete sources within the UHE HAWC source extent, providing upper limits on local enhancements in the FIR densities. For J1825$-$134, 
an enhancement factor of up to 16 is possible for all IRAS wavelengths.
An enhancement factor of 4 is consistent with the data for the region around J2019+368. For J1907+063, no enhancement due to FIR sources within the angular extent of the source in the range \SIlist{60;100}{\micro\meter} is possible. Enhancements due to more distant emission regions (within the Galactic Plane) remain possible, but are not essential for IC models of this object (see later, and Appendix C).

The associated pulsars have characteristic spin-down ages $P/2\dot{P}\approx17-21$ kyr and powers $L_{36}=2.8-3.4$ \citep{HAWC_UHE}.
As pulsars are known to accelerate electrons beyond \SI{100}{\tera\electronvolt} \citep{CrabPeV} and meet the requirement implied by Figure~\ref{fig:hillas_and_gammagamma_limits}, the association favours an IC interpretation. As mentioned before, the lack of a clear correlation with nearby target material further supports this explanation.
The inferred system sizes given in \citet{HAWC_UHE} are in the range $6-22$ pc, likely determined by the combination of radial diffusion and losses. Age limited diffusion can not be conclusively ruled out, but the implied radiative inefficiency would place severe demands on the electron injection power above 10 TeV (see below).
All sources fulfil
the $\gamma\gamma$-absorption constraints of Figure \ref{fig:hillas_and_gammagamma_limits} for $T =\,$\SI{50}{\kelvin} and $B$-fields up to \SI{10}{\micro\gauss}.
For the multi-component photon fields used in our modelling (see below), detailed calculations show that all sources have intrinsic $\gamma\gamma$-absorption less than \SI{0.25}{\percent}, which can be safely neglected.

\begin{figure}
    \includegraphics[width=0.485\textwidth]{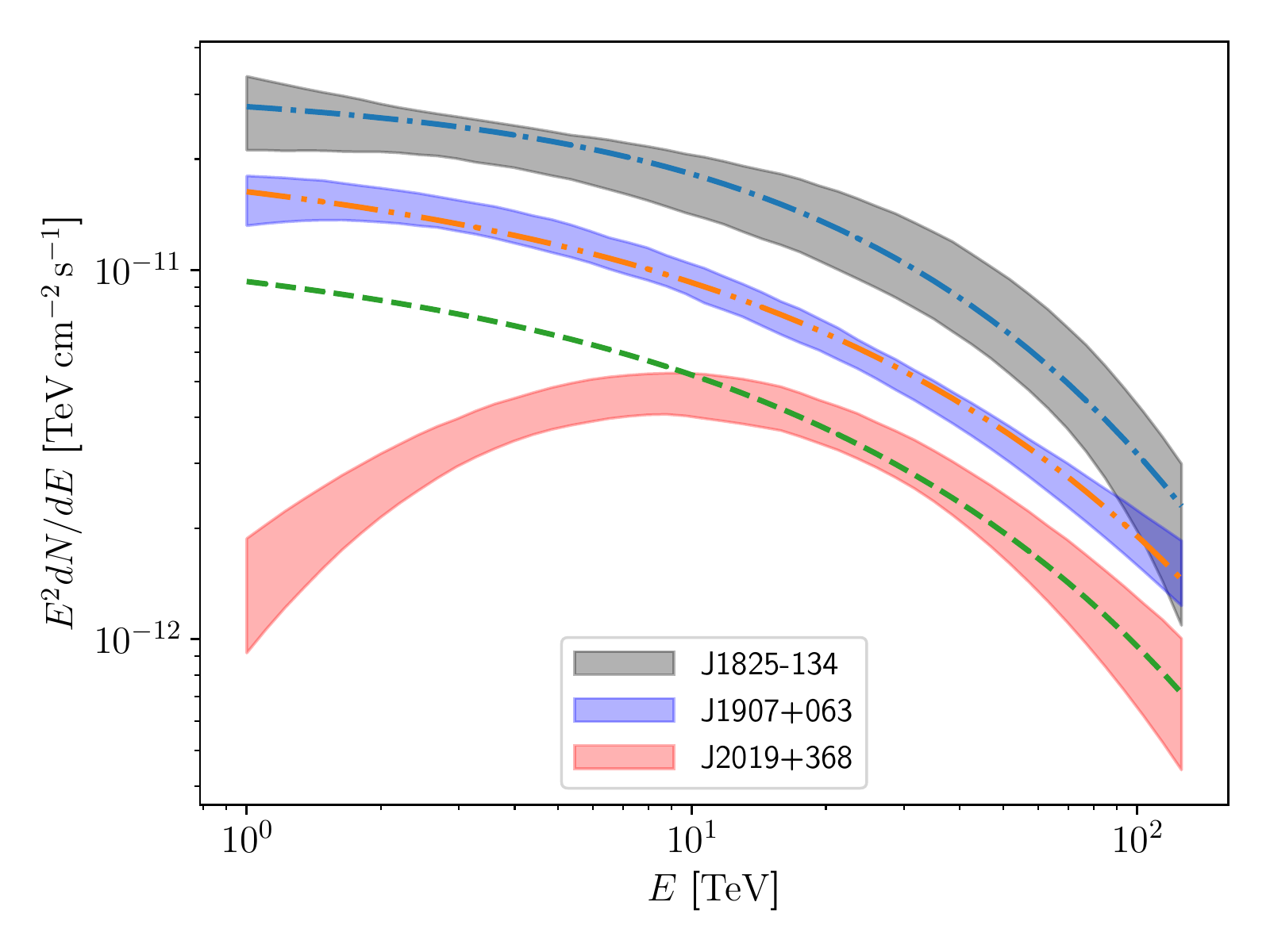}
    \caption{Comparison of the measured spectra of the three UHE HAWC sources to representative equilibrium IC emission models. The error bands of the source spectra are the systematic errors extracted from Figure 3 in \citet{HAWC_UHE} and the statistical error from the best fit. All IC model curves were obtained with injection index $\alpha$=2, $B=$\SI{3}{\micro\gauss} and a photon field of the CMB plus the model of \citet{Popescu2017} multiplied by an enhancement factor $\eta$. For the three sources we adopt $\eta=3$ ($\Xi_{\rm IC} = 42$), $E_{\text{cut}} = \SI{350}{\tera\electronvolt}$ (J1825$-$134 / PSR~J1826$-$1334), $\eta=1.0$ ($\Xi_{\rm IC} = 8$), $E_{\text{cut}} = \SI{480}{\tera\electronvolt}$, (J1907+063) and $\eta=2$ ($\Xi_{\rm IC} = 9$), $E_{\text{cut}} = \SI{400}{\tera\electronvolt}$, (J2019+368). $\Xi_{\rm IC}$ includes the total target photon spectrum. 
    }
    \label{fig:hawc}
\end{figure}

Figure~\ref{fig:hawc} compares the spectra of UHE HAWC sources to representative IC models with $B=\SI{3}{\micro\gauss}$ and a photon target of CMB plus the axisymmetric model \citep{Popescu2017} at the location of the sources, enhanced by a factor $\eta$. The equilibrium electron spectra were calculated for an exponential cutoff power-law injection spectrum with index $\alpha$, cutoff energy $E_{\text{cut}}$ and some fraction of the pulsar spin down power injected into electrons above  \SI{1}{\giga\electronvolt}.
As we are interested in equilibrium spectra, we focus only on the spectrum near the cutoff. The spectra of J1825$-$134 and J1907+063 can be reproduced, although the model spectrum for J2019+368 is consistent only above \SI{10}{\tera\electronvolt}. This is naturally explained if the electrons emitting below \SI{10}{\tera\electronvolt} are not yet in equilibrium due to the youth of the source  \cite[see e.g.][]{VikasThesis}. To obtain the required flux levels, we need to inject between \SI{\sim 1}{\percent} (J2019+368) and \SI{\sim 13}{\percent} (J1825$-$134) of the spin down power of the pulsars into electrons above \SI{10}{\tera\electronvolt}.
For J1825$-$134 there are two possible pulsar associations \citep{HAWC_UHE}. This leads to different radiation energy densities in the model of \citet{Popescu2017}; we require an enhancement factors of $\eta=3$ for association to PSR\,J1826$-$1334, and $\eta=5$ ($\Xi_{\rm IC} = 34$) for PSR\,J1826$-$1256. The resulting spectrum is similar. J1907+063 and J2019+368 were modelled with $\eta=1$ and $\eta=2$ respectively.

All models are consistent with the upper limits on local source contributions derived from IRAS maps and the spin-down power of the corresponding pulsars. 
Note however that the parameters used for model fits are not unique. Either a larger value of $\eta$ or lower value of $B$ would shift the cutoff in the $\gamma$-ray spectrum to higher energies, an effect compensated by a smaller value of $E_{\text{cut}}$. Lower values of $\eta$ would have the opposite effect, but a different value of $\alpha$ will mask the difference. The spectrum of J1907+063, for example, is comparable with $\eta=0.5$ or 1.5 and the spectrum of J2019+368 with $\eta = 1$. The models nevertheless illustrate that IC scenarios with plausible parameters exist for these sources. 

Multi-wavelength observations can help to disentangle the origin of the radiation. In the energy range of the Fermi-LAT telescope, more sophisticated models are necessary, since the spectra will not be in equilibrium. However, the same high energy electrons producing emission above \SI{10}{\tera\electronvolt} will also produce synchrotron emission in the X-ray band. 
The synchrotron flux from X-ray emitting electrons will be approximately $\Xi_{\rm IC}$ times smaller than their IC flux. This is broadly consistent with existing X-ray data \citep{Mizuno2017}, although a detailed source-specific investigation beyond the scope of this Letter is needed.

\section{Conclusions}\label{sec:05}
As shown in this Letter, hard IC spectra up to and beyond 100 TeV are possible wherever IC losses dominate over synchrotron losses for sufficiently high energies. These conditions are likely not met in average large-scale environments of the Milky Way. High latitudes and large galactocentric radii are potential exceptions, although sources are scarce at such locations. However, the sites of PeV electron acceleration are likely to sample significantly higher than galactocentric average radiation energy densities, located in spiral arms and in general 
associated with high-mass star formation; regions of intense far-infrared emission. In such environments hard IC emission is possible if the region is able to confine the high-energy particles long enough, but also small enough to prevent strong $\gamma\gamma$-absorption. The recently detected HAWC observatory UHE sources at \SI{100}{\tera\electronvolt} \citep{HAWC_UHE} support such leptonic scenarios, with energy dissipation of the pulsar wind, for example via shocks, as the favoured accelerator.
As pulsars are the only class of galactic sources known to produce $>100$ TeV electrons, and no theory currently existing can account for very hard ($\alpha  \approx 1$) spectra to these energies in pulsars, the combination of a powerful pulsar in an IR photon dominated environment, appear to be essential features of any leptonic model.
We stress that none of the HAWC sources exhibit a correlation with tracers of target material, which remains an obstacle for hadronic scenarios.
A population of UHE sources with well characterised spectra are important to establish the contribution of leptonic processes at UHE. The LHAASO observatory, which recently started operation, and the future CTA and SWGO observatories offer tremendous promise in this regard.

\acknowledgments
We thank the anonymous referee for their valuable suggestions and comments. The GAMERA code is publicly available at \url{https://github.com/libgamera/GAMERA}.

\bibliographystyle{aasjournal}

\appendix

\section{Equilibrium solution for radial dependent cooling rate}\label{sec:AppendixC}

Here we explore the evolution of a synchrotron cooling spectrum assuming particles escape the acceleration site at the local cooling limit. We consider only the case in which the field is strong, in the sense that the magnetic energy density in the immediate vicinity of the accelerator greatly exceeds that on the scales relevant to the observed $\gamma$-rays. If particles are frozen into the radial flow, adiabatic losses will be detrimental. We thus ignore advection and consider the diffusion limit. 

Assuming scattering maintains a near isotropic distribution locally, the differential number of particles $N(E,r)$ satisfies  
\begin{equation}
\pdiff{N}{t} + \frac{1}{4 \pi r^2}\pdiff{j}{r}+
\pdiff{}{E}\left[ \dot{E}_{\rm cool}~ N\right] = 0
\end{equation}
where 
\begin{equation}
j(r,E)= -4 \pi r^2 D \pdiff{N}{r}
\end{equation}
is the diffusive radial flux, and $\dot{E}_{\rm cool}(E,r)$ the cooling rate which we take to be dominated by synchrotron radiation in a radially decreasing magnetic field $B\propto 1/r$, i.e. $\dot{E}_{\rm cool} \propto E^2/r^2$.

With these assumptions, we seek steady-state solutions allowing for continuous injection of electrons/positrons across the inner boundary corresponding to a spherical surface of radius $r=r_0$. This is best given in terms of the diffusive flux, which we take as a single power-law with exponential cut-off:
\begin{equation}
j(r_0,E) =  Q_0E^{-\alpha}e^{-E/E^*}\, .
\end{equation}
In the synchrotron loss limited picture of relativistic shock acceleration, a theoretical upper limit to $E^*$ can be determined by equating the electron gyro-time to its cooling time \citep{GiacintiKirk}. This gives 
\begin{equation*}
    E_{\rm max,sh} = \sqrt{\frac{6 \pi e}{\sigma_T B}}m_{\rm e} c^2
\end{equation*}
which, on assuming the magnetic flux through the (shock) surface at radius $r_0$ to be $4 \pi r_0^2 U_B c = \eta_{\rm eq} L_{\rm SD}$, with $U_B$ the magnetic energy density, $L_{\rm SD}$ the spin-down power and $\eta_{eq}<1$ an equipartition factor can be expressed numerically as
\begin{equation*}
    E_{\rm max,sh} \approx 10 \left(\frac{r_0}{0.1~{\rm pc}}
\right)^{1/2} \left(\frac{\eta_{eq}L_{\rm SD}}{10^{36}{\rm erg ~s}^{-1}}\right)^{-1/4}  {\rm PeV} 
\end{equation*}
On the other hand, applying the same analysis to the Hillas limit, with scattering velocity $\beta\approx 1$, one finds 
\begin{equation*}
    E_{\rm Hillas} = e \beta B r_0 \approx 2.5 \left(\frac{\eta_{eq}L_{\rm SD}}{10^{36}{\rm erg ~s}^{-1}}\right)^{1/2} ~{\rm PeV}  
\end{equation*}
Both are consistent with PeV electron production.

Physical solutions to the transport equation require $N(r=\infty)=0$.
The method of solution follows that of \cite{WebsterLongair}, generalized here to include energy dependent diffusion, $D = D_0 E^{\delta}$. 

Subject to the constraint $0\leq \delta < 1$, the solution is

	\begin{equation}
	j(r,E)=   \frac{Q_0}{ \sqrt{\pi}} E^{-\alpha} \frac{r}{r_0}   
	\int_{z_0}^\infty \left(1-\frac{z_0^2}{z^2}\right)^{\frac{\delta+\alpha-2}{1-\delta}}
	\exp\left[-\frac{E}{E^*}\left(1-\frac{z_0^2}{z^2}\right)^{\frac{1}{\delta-1}}\right]\exp\left[-\left(\frac{\ln(r/r_0) + z^2}{2z}\right)^2\right]  ~dz
	\end{equation}

where 
\begin{equation*}
z_0=\sqrt{ 
	\frac{(1-\delta)r_0^2}{D\,t_{\rm cool,0}} } 
\ln\left(\frac{r}{r_0}\right) \mbox{~~~with~~~} t_{\rm cool,0} = \left. \frac{E}{\dot{E}_{\rm cool}}\right|_{r=r_0}\enspace.  
\end{equation*}
In the limit of $\delta, 1/E^* \rightarrow 0$, this is equivalent to the result previously found by \cite{WebsterLongair}. If we additionally take $\alpha=2$, in this limit, the integration can be carried out, giving 
\begin{equation}
j(r,E)=  \frac{Q_0}{2} E^{-2} \left[ 
{\rm erfc} (Z_-) + \frac{r}{r_0} {\rm erfc} (Z_+)\right]  \nonumber
\end{equation}
where ${\rm erfc}(x)$ is the complementary error function, and
\begin{equation}
Z_\pm=\frac{\sqrt{D\,t_{\rm cool,0}}}{2r_0}\left[
\frac{r_0^2 \ln(r/r_0)}{D\,t_{\rm cool,0}}   \pm 1
\right]\nonumber
\end{equation}
For the cases of interest, the term with $Z_-$ dominates, and thus a break occurs at $Z_- \approx 1$ i.e. at a break energy of
\begin{align}
E_{\rm br} &\approx \frac{3 m_e^2 c^3 D}{ 4 \pi r_0^2 \sigma_{\rm T} U_B(r_0)}
\left[\frac{1+\sqrt{1+\ln(r/r_0)}}{\ln(r/r_0)} \right]^2  \\
&\approx 0.5 \left(\frac{D}{10^{26}~{\rm cm^2 ~s}^{-1}}
\right) \left(\frac{\eta_{eq}L_{\rm SD}}{10^{36}{\rm erg ~s}^{-1}}\right)^{-1}  {\rm PeV} \nonumber
\end{align}

Hence, for a pulsar with spin-down power of $L_{\rm SD} \gtrsim 10^{36}~{\rm erg ~s}^{-1} $, and applying the standard limits of \citet{Hillas1984}, it is apparent that electrons at energies $\gg 100$~TeV can survive to large radii. The dependence on the position of the acceleration site is weak. Provided $D(E>100 ~{\rm TeV}) \gg 10^{26}~{\rm cm^2 ~s}^{-1} $, the steady-state flux at large radii is insensitive to the scattering rate. 
Such diffusion parameters are consistent with the HAWC sources, which restrict the allowed values to $D\ll 10^{28} (R_{\rm system}/{\rm pc})~{\rm cm}^2 ~s^{-1}$ at the observed energies (i.e. diffusion time should exceed the light crossing time).

\section{Changes in the resulting $\gamma$-ray spectrum for the radiation field models used in Figure 4\label{sec:AppendixA}}

\begin{figure*}[!h]
    \centering
    \includegraphics[width=0.45\textwidth]{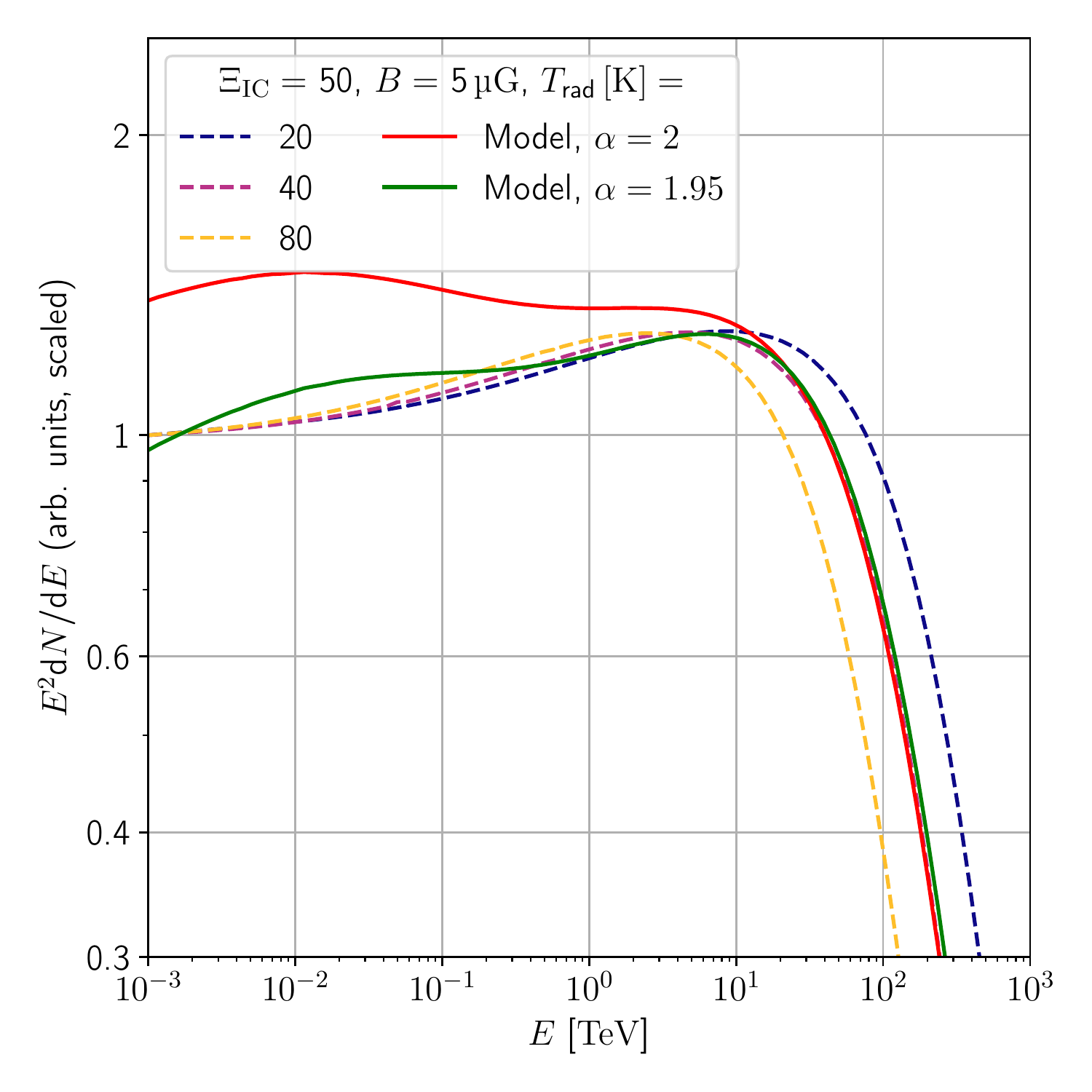}
    \includegraphics[width=0.45\textwidth]{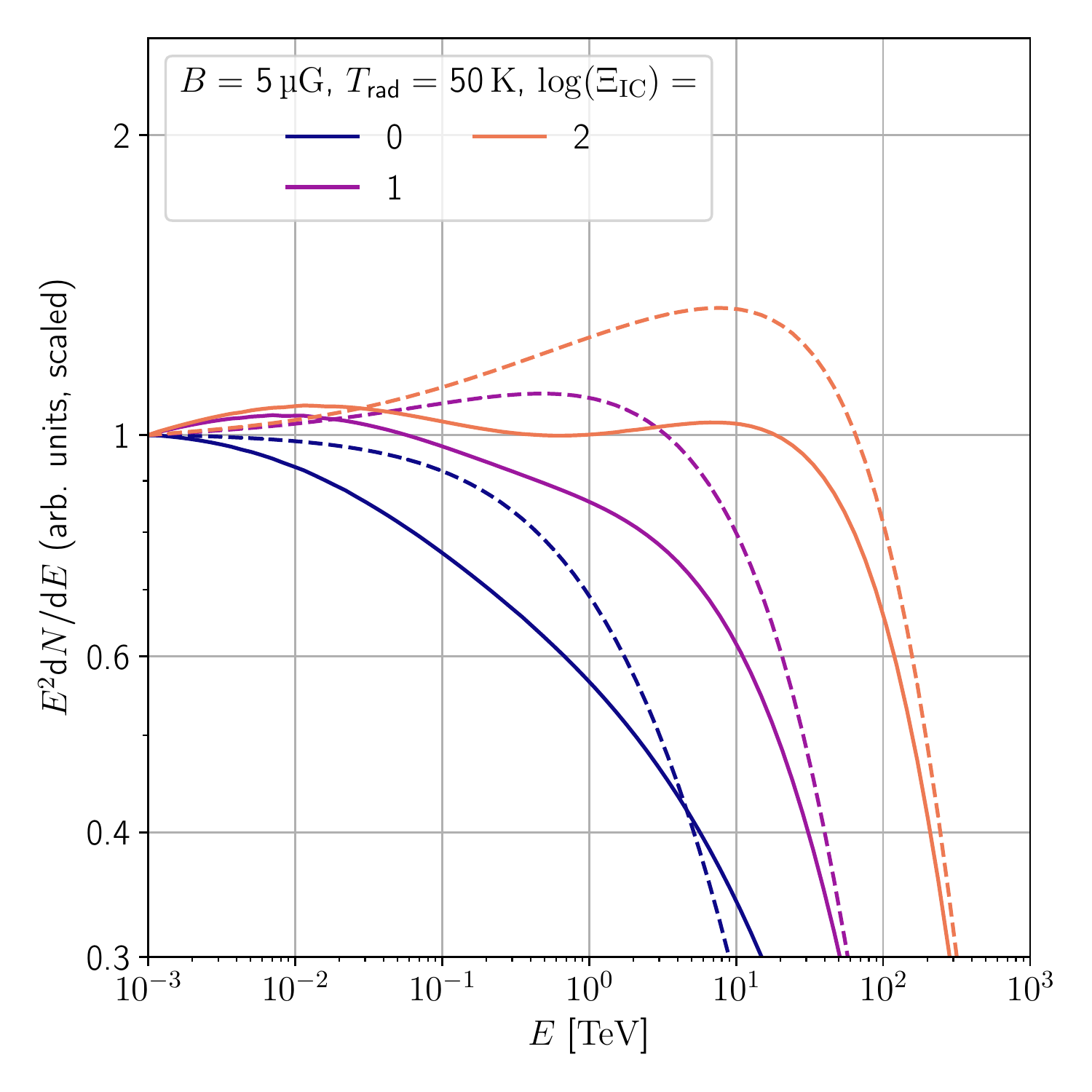}
    \caption{{Comparison of the $\gamma$-ray spectra resulting from realistic galactic radiation fields used to model the HAWC sources in Figure \ref{fig:hawc} with $\gamma$-ray spectra resulting from single black body fields. We used the dust and starlight model from \cite{Popescu2017} at the assumed location of the HAWC source J1907+063 scaled such that the total energy density in the dust radiation accounts for the ratio $\Xi_{\rm IC}$ and added the CMB. The left panel shows curves for $\Xi_{\rm IC} = 50$. The dashed lines show the cases of single black bodies with different temperatures and an injection index for the electrons of $\alpha = 2$. The solid lines show the results from the realistic photon field for $\alpha = 2$ and $\alpha = 1.95$. In the right panel we compare $\gamma$-ray spectra for different values of $\Xi_{\rm IC}$ from a black body of $T=\SI{50}{\kelvin}$ (dashed lines) with the corresponding spectra for the same values of $\Xi_{\rm IC}$ with the realistic photon field (solid lines). In both figures, the $B$-field was fixed at \SI{5}{\micro\gauss}.}}
    \label{fig:realistic_field_comparison}
\end{figure*}

The radiation fields used to model the HAWC sources in Figure \ref{fig:hawc} consist of three different components: direct emission from stars, emission from dust, and the CMB. Figure \ref{fig:realistic_field_comparison} shows a comparison between the resulting $\gamma$-ray spectra and that derived from a pure black-body spectrum for a range of temperatures. Since the photon spectra used at the different HAWC source locations, apart from the modest enhancement factor are very similar, we selected one of the fields (at the location of J1907+063), re-scaled such that the total emission of the dust equates to the value of $\Xi_{\rm IC}$. The left panel of Figure \ref{fig:realistic_field_comparison} compares the the black body spectra of different temperatures (dashed lines) with the multi-component model (solid lines) for fixed
$B=\SI{5}{\micro\gauss}$ and $\Xi_{\rm IC} = 50$. The powerlaw injection index of the electron spectrum was $\alpha = 2$ for the black body curves and the red solid line, but $\alpha = 1.95$ for the green solid line. The spectra are normalised to facilitate the comparison of the shapes. Whereas the black body $\gamma$-ray spectra (and all spectra in the right panel) are normalised in the same way as in Figure \ref{fig:break} such that the value at \SI{1}{\giga\electronvolt} is unity, the green and red curves were scaled slightly differently to be able to compare the cutoff energies better. We first concentrate on the comparison for the same $\alpha$ value (red solid line). The high-energy cutoff due to synchrotron cooling is evidently close to that of the black body case with $T = \SI{40}{\kelvin}$. Relative to the black body cases, the low energy spectrum is different: the spectrum being slightly harder at the lowest energies shown, but for the chosen parameters softens above \SI{\sim 20}{\giga\electronvolt}. This follows from the Klein-Nishina suppression of star-light scattering at higher energies. The effect can be compensated by a slightly harder electron injection spectrum index, which is shown with the green solid line, where $\alpha = 1.95$. The shape above \SI{\sim 300}{\giga\electronvolt} is nearly identical to the one for $T = \SI{40}{\kelvin}$.

{For low ratios or low $B$-fields, the CMB starts to become important and change the spectral shape at higher energies. This effect is shown in the right panel of Figure \ref{fig:realistic_field_comparison}, where we compare the equilibrium spectra for different values of $\Xi_{\rm IC}$ of the realistic photon field (solid lines) with the black body case (dashed lines). If $\Xi_{\rm IC}$ is low, the spectrum of the realistic photon field decreases less rapidly and remains harder until higher energies because of the influence of the CMB.}

\section{ Local energy densities at the location of the HAWC sources}
We use the axi-symmetric radiation field model from
\citet{Popescu2017} to estimate the energy densities
of the diffuse interstellar radiation field (ISRF)
in the FIR at the locations of the HAWC sources. These values
are tabulated in Table \ref{tab:urad_popescu}, and correspond to the assumption
that the sources are at the same distance from the
Sun as the associated pulsars. We note the axisymmetric model
of Popescu et al. will underestimate the diffuse ISRF in
regions of enhanced volume emissivity, such as spiral arms,
in spectral domains where the emission is optically thin.
As discussed in the main text, in such cases,
one expects the in situ ISRF to be higher
than the values in Table \ref{tab:urad_popescu} by a factor
of a few, corresponding to the typical density contrast
between the arm and inter-arm regions
\cite[see also the discussion in][]{Popescu2017}.

\begin{table}[]
	\centering
	\begin{tabular}{c|c |c| c| c}
		\multicolumn{1}{c|}{Name}& 
		\multicolumn{4}{c}{$u_{\text{ISRF}}$ [\si{\electronvolt\per\cubic\centi\meter}]} \\ \hline
		&    $12\mu$m  & $25\mu$m  & $60\mu$m  & $100\mu$m \\ \hline
		J1825$-$134, \SI{3.61}{\kilo\pc}  & 0.17 & 0.15 &  0.37 & 0.67 \\ \hline
		J1825$-$134, \SI{1.55}{\kilo\pc} & 0.09  & 0.06 &  0.13 & 0.31 \\ \hline
		J1907+063 & 0.09  & 0.06 &  0.14 & 0.32 \\ \hline
		J2019+368 & 0.06  & 0.04 &  0.07 & 0.18 \\
	\end{tabular}
	\caption{Energy densities at the assumed locations of the HAWC sources at different wavelengths from the model of \cite{Popescu2017}. For J1825$-$134, there are two associations to pulsars: PSR1826$-$1334 at a distance of \SI{3.61}{\kilo\pc} from the Sun, and PSR1826$-$1256 at \SI{1.55}{\kilo\pc}.}
	\label{tab:urad_popescu}
\end{table}

Significant additional radiation fields over and above
the diffuse ISRF are expected in the vicinity of discrete sources.
To estimate these potential enhancements, we examined FIR images around the locations of the HAWC \SI{100}{\tera\electronvolt} sources. We use the IRAS survey for this purpose, as it's spectral grasp covers the \SIrange{25}{100}{\micro\meter} region
where dust emission SEDs of star forming regions typically peak, and is
sensitive to structure on the angular scale of the resolved HAWC sources.
The resulting upper limits on additional local contributions
to the radiation fields from discrete sources, corresponding
to the assumption that these sources are at the same distance as the
gamma-ray sources, are given in Table \ref{tab:urad_iras}.

\begin{figure}
    \centering
    \includegraphics[width=0.7\textwidth]{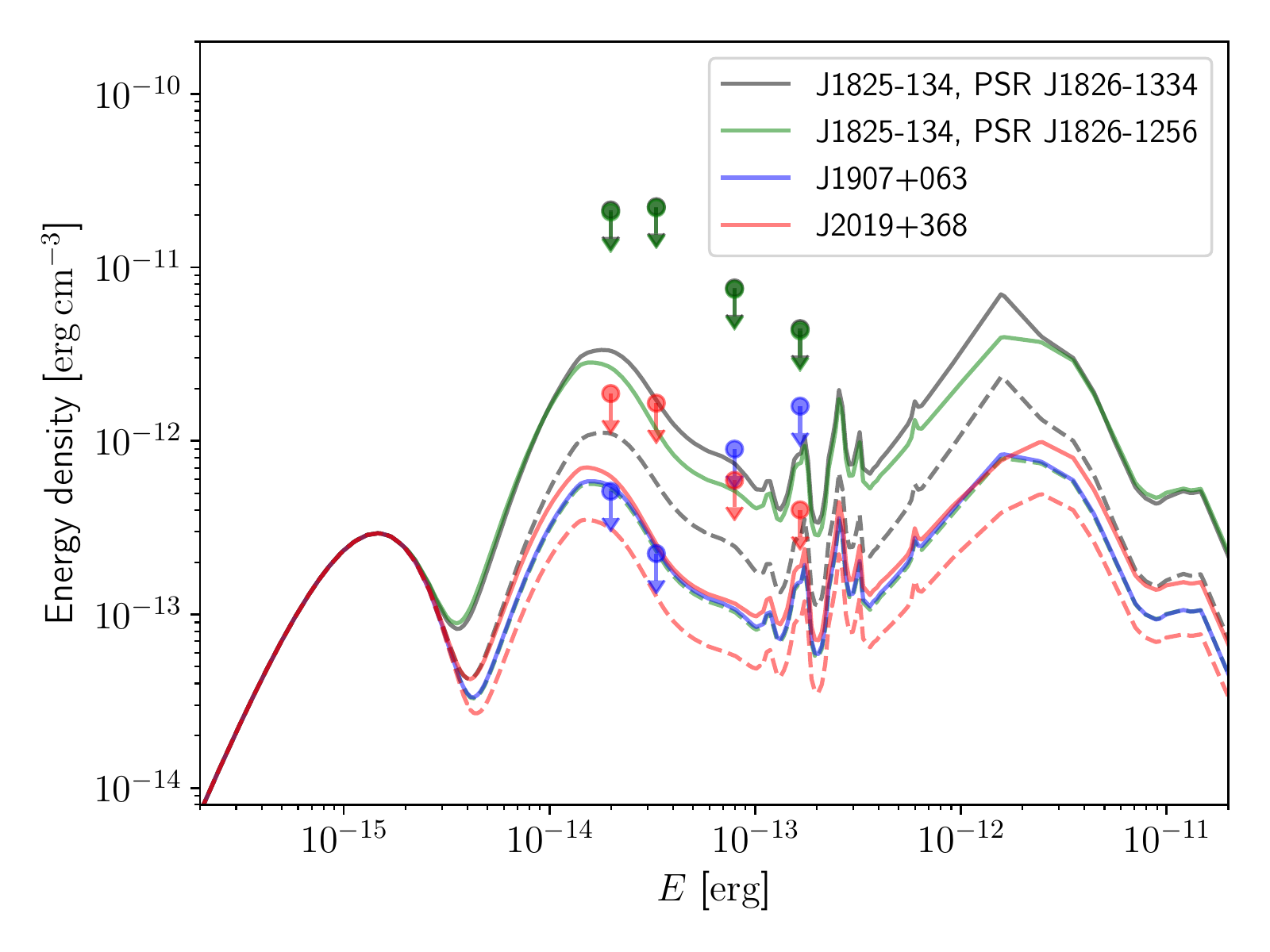}
    \caption{{Photon spectra used to model the HAWC sources (see Section \ref{sec:04}). The different colours represent the different models for the sources, for J1825-134 there are two models due to the two possible pulsar associations. Dashed lines show the level of emission from the model of \cite{Popescu2017}, solid lines the enhanced radiation. The upper limits show the maximum radiation densities possible at the source positions derived from IRAS maps (the values in Table \ref{tab:urad_iras} added to the values in Table \ref{tab:urad_popescu}) in the respective colours. The radiation fields are consistent with these values.}}
    \label{fig:photon_fields}
\end{figure}

{In Figure \ref{fig:photon_fields} we show the derived upper limits on the total radiation energy density, which are the values in Table \ref{tab:urad_iras} added onto the values of the background field shown in Table \ref{tab:urad_popescu}. We also plotted for each model in the same colours the radiation fields used. The dashed lines are the values of the background field from the model of \cite{Popescu2017} at the respective locations, and the solid lines show the enhanced radiation densities used. The bump at $\sim \SI{e-15}{\erg}$ is the CMB. For J1907+063 no dashed line is shown, because no enhancement was used. The upper limits for J1825-134 for both associated pulsar locations are not distinguishable, since the upper limits from the IRAS maps are much larger than the background fluxes. All upper limits are at the same level or above the respective fluxes of the models.
}

\begin{table}[]
	\centering
	\begin{tabular}{c|c |c| c| c}
		\multicolumn{1}{c|}{Name}& 
		\multicolumn{4}{c}{$u_{\text{discrete}}$ [\si{\electronvolt\per\cubic\centi\meter}]} \\ \hline
		&    $12\mu$m  & $25\mu$m  & $60\mu$m  & $100\mu$m \\ \hline
		J1825$-$134 & $< 2.6$  & $< 4.6$  & $< 13.6$  & $< 12.7$ \\ \hline
		J1907+063 & $< 0.9$ & $< 0.5$ & $0.0$ & $0.0$ \\ \hline
		J2019+368 & $< 0.19$ & $< 0.33$ & $< 0.96$  & $< 0.99$  \\
	\end{tabular}
	\caption{Upper limits on additional contributions to the energy densities at the locations of the pulsars associated with the HAWC sources deduced from IRAS data. These limits are independent of the distance towards the sources.}
	\label{tab:urad_iras}
\end{table}

\end{document}